\documentclass{elsart}

\usepackage{graphicx}

\usepackage{amssymb}

\begin{document}

\begin{frontmatter}


\title{Numerical integration methods for large-scale biophysical simulations}

\author[TS]{Edoardo Milotti\corauthref{cor1}}
\ead{milotti@ts.infn.it}
\author[TS]{Alessio Del Fabbro}
\ead{Alessio.Delfabbro@ts.infn.it}
\author[VR]{Roberto Chignola}
\ead{chignola@ts.infn.it}
\corauth[cor1]{corresponding author: Edoardo Milotti}
\address[TS]{Dipartimento di Fisica dell'Universit\`a di Trieste, and I.N.F.N. -- Sezione di Trieste, Via Valerio 2 -- I-34127 Trieste, Italy }
\address[VR]{Dipartimento di Biotecnologie dell'Universit\`a di Verona, and I.N.F.N. -- Sezione di Trieste, Strada Le Grazie 15 - CV1, I-37134 Verona, Italy
}




\begin{abstract}
Simulations of biophysical systems inevitably include steps that correspond to time integrations of ordinary differential equations. These equations are often related to enzyme action in the synthesis and destruction of molecular species, and in the regulation of transport of molecules into and out of the cell or cellular compartments. Enzyme action is almost invariably modeled with the quasi-steady-state Michaelis-Menten formula or its close relative, the Hill formula: this description leads to systems of equations that may be stiff and hard to integrate, and poses unusual computational challenges in simulations where a smooth evolution is interrupted by the discrete events that mark the cells' lives. This is the case of a numerical model (Virtual Biophysics Lab - VBL) that we are developing to simulate the growth of three-dimensional tumor cell aggregates (spheroids). The program must be robust and stable, and  must be able to accept frequent changes in the underlying theoretical model: here we study the applicability of known integration methods to this unusual context and we describe the results of numerical tests in situations similar to those found in actual simulations. 
\end{abstract}

\end{frontmatter}

\section{Introduction}
\label{intro}
Most mathematical models of biochemical and biophysical processes in cells are described by nonlinear differential systems that cannot be handled with analytical methods and require numerical integrations \cite{CB,sysbio}. However the high speed of computers and the sophisticated computational methods that are available today are powerful tools that allow the numerical exploration of these exceedingly interesting dynamical systems. This also suggests that eventually the biophysical models will no longer be analytic, but mostly computational; and indeed we are now developing a program that simulates tumor spheroids (VBL, Virtual Biophysics Lab) \cite{VBL0,VBL1,VBL2,VBL3}, which includes a reduced Ð but still quite complex Ð description of the biochemistry of individual cells, plus many diffusion processes that bring oxygen and nutrients into cells and metabolites into the environment. 

In a numerical model like VBL, each cell is described by a reduced metabolic network, and by other mechanisms that include both discrete deterministic  and stochastic events. The description is thus mixed, with smooth evolutions interspersed with discrete steps. The exchange of molecules with the surrounding environment means that transport into and out of cells is closely linked with diffusion processes that involve the whole cluster of cells, and finally lead to a very large set of (time) differential equations. 
Simulation steps between discrete events require the integration of nonlinear differential equations that describe the individual cell's clockwork, and the integration of the diffusion equations. These integrations are carried out under widely different conditions, in a changing environment, and for this reason they need algorithms that are both unconditionally stable and free from unwanted artifacts. These conditions are not always fulfilled in the existing literature, and we feel that a detailed understanding of the underlying mathematical and computational bases may be important not just for us, but for other workers in the field of systems biology as well.

Eventually the whole simulation program must run smoothly, and it must be free of stability problems. Thus it is very important to ensure that step by step integrations are stable and do not bring the biochemical variables into unphysical regions (e.g., no concentration must ever become negative).

\section{Biophysical models}
\label{bio}

Biophysical models are not arbitrary dynamical systems: indeed the phenomena of life are characterized by remarkable stability properties, as most of them display either homeostasis or (nearly) stable limit cycles. At the basic reaction level, many processes obey a straightforward enzyme kinetics, regulated by the well-known Michaelis-Menten equation \cite{CB}, or by a variant that applies to cooperative processes, the Hill equation \cite{weiss}. 

Here we start with a special case that displays generic features shared by many other processes, namely the transport of glucose into and out of cells and its conversion into glucose-6-phosphate (G6P; this is part of the reduced metabolic model incorporated in our simulator of cell metabolism, growth and proliferation \cite{VBL1,VBL2}). We begin with the equations for a single cell in a stable environment: 
\begin{eqnarray}
\nonumber
 V_{in} \frac{{d\left[ {G_{in} } \right]}}{{dt}} &=& \frac{{v_{\max ,1} \left[ {G_{extra} } \right]}}{{K_1  + \left[ {G_{extra} } \right]}} - \frac{{v_{\max ,1} \left[ {G_{in} } \right]}}{{K_1  + \left[ {G_{in} } \right]}} \\
 \nonumber
 &&- \frac{{v_{\max ,2} \left[ {G_{in} } \right]^2 }}{{\left( {K_2  + \left[ {G_{in} } \right]} \right)\left( {K_a  + \left[ {G_{in} } \right]} \right)}} \\
  \label{conc1}
 &&- \frac{{v_{\max ,22} \left[ {G_{in} } \right]^2 }}{{\left( {K_{22}  + \left[ {G_{in} } \right]} \right)\left( {K_a  + \left[ {G_{in} } \right]} \right)}} \\ 
 \nonumber
V_{extra} \frac{{d\left[ {G_{extra} } \right]}}{{dt}} &=&  - \frac{{v_{\max ,1} \left[ {G_{extra} } \right]}}{{K_1  + \left[ {G_{extra} } \right]}} 
 + \frac{{v_{\max ,1} \left[ {G_{in} } \right]}}{{K_1  + \left[ {G_{in} } \right]}}\\
 \label{conc2}
&&+ D\frac{{\left[ {G_{env} } \right] - \left[ {G_{extra} } \right]}}{\Delta }S
\end{eqnarray}
Here the dynamical variables are the glucose concentration inside the cell $\left[ {G_{in} } \right]$, and the concentration in the extracellular space $\left[ {G_{extra} } \right]$; the environmental glucose concentration $\left[ {G_{env} } \right]$ is fixed and is one of the model parameters. The introduction of the extracellular space may  look like a useless nuisance in this case, but it is justified on two different, and both important, grounds. The extracellular space is absolutely needed in multicellular systems, because in that case it becomes the scaffolding that allows the diffusion of many substances in closely bound cell clusters (and corresponds to an actual biological entity, the free space available in the extracellular matrix \cite{jain,krol,jin}), and a consistent treatment requires its extension also towards the environment; secondly, the extracellular space, even in this simple case, acts as the cell-environment  interface, and corresponds to the buffer region actually observed in diffusion measurements in tumor spheroids \cite{GMK}. 

We assume a spherical cell of radius $r \approx 5 \mu\mathrm{m}$, so that the surface area is $S = 4\pi r^2 \approx 3\cdot 10^{-10} \mathrm{m}^2$, and the volume is $V_{in} =(4/3)\pi r^3 \approx 5\cdot 10^{-16} \mathrm{m}^3$. The first two terms on the rhs of equation (\ref{conc1}) correspond to the transport of glucose from the extracellular space into the cell and the reverse process of transport from the inside of the cell back to the extracellular space (enacted by the reversible GLUT transporters, see \cite{MOw,air} and references therein); since this transport process is facilitated \cite{MOw,air} we describe it with two Michaelis-Menten terms, where the maximum transport speed $v_{\max ,1}$ is proportional to the cell's surface \cite{VBL1,MOw,air}, and $K_1$ is an experimental parameter \cite{VBL1}.The other two terms correspond to a cascade of two Michaelis-Menten processes that involve the enzymes glucokinase and hexokinase, and is regulated by parameters that do not depend on cell size \cite{heim}.

The second equation (eq.  (\ref{conc2})), in addition to the GLUT-mediated transport terms, includes a standard diffusion term that corresponds to the diffusion of glucose from the surrounding environment into the extracellular space. This term is a discretization of glucose flux across the surface $S$ (which is approximately the area of the cell-extracellular space interface); the distance used to compute the concentration gradient is the thickness of the extracellular layer, $\Delta \approx 0.2 \mu\mathrm{m}$. 

Although equations like (\ref{conc1}) and (\ref{conc2}) are usually cast in terms of concentrations, we prefer to use masses, as this makes total mass conservation stand out clearly: 
\begin{eqnarray}
\nonumber
 \frac{{dm_{in} }}{{dt}} &=& \frac{{v_{\max ,1} m_{extra} }}{{V_{extra} K_1  + m_{extra} }} 
- \frac{{v_{\max ,1} m_{in} }}{{V_{in} K_1  + m_{in} }} \\
\nonumber
&&- \frac{{v_{\max ,2} m_{in} ^2 }}{{\left( {V_{in} K_2  + m_{in} } \right)\left( {V_{in} K_a  + m_{in} } \right)}} \\
\label{mass1}
&& - \frac{{v_{\max ,22} m_{in} ^2 }}{{\left( {V_{in} K_{22}  + m_{in} } \right)\left( {V_{in} K_a  + m_{in} } \right)}} \\ 
\nonumber
 \frac{{dm_{extra} }}{{dt}} &=&  - \frac{{v_{\max ,1} m_{extra} }}{{V_{extra} K_1  + m_{extra} }} + \frac{{v_{\max ,1} m_{in} }}{{V_{in} K_1  + m_{in} }} \\
 \label{mass2}
 && + \frac{{DS}}{\Delta }\left( {\left[ {G_{env} } \right] - \frac{{m_{extra} }}{{V_{extra} }}} \right) 
\end{eqnarray}
We postpone the listing of model parameters to section \ref{res}, where we summarize the results of numerical tests.

\section{Integration methods}
\label{numint}

There is a broad and comprehensive literature on integration methods, and at first it may seem that in large-scale biophysical simulations the choice is only limited by the required accuracy and by the available computing power, however it is not so. In a simulation like VBL \cite{VBL1,VBL2} there are many cells and the number of equations is quite large: our aim is to simulate at least one million cells, which corresponds to a spherical cell cluster with a diameter of one millimeter. In VBL we make several simplifying assumptions and one of them is that each cell has its own extracellular space: this means that in such a system -- for each molecular species which diffuses in the cell cluster --  there are about two million local concentration variables (for each cell there is an intracellular and an extracellular concentration) and a corresponding number of equations. 

The simulation includes random events as well (such as the partitioning of organelles at the time of cell division), so that numerical inaccuracies are absorbed in this much larger randomness and turn out to be much less important than they are in strictly deterministic systems. 

Moreover, the life of cells is marked by discrete events: this discreteness produces endless transient responses in the deterministic part of the simulation algorithm, and the most evident discrete process -- cell division -- also changes the number of variables and equations. 

Last but not least, there is a wide range of volumes (cells and extracellular spaces), and this leads to very stiff systems of equations: in fact while the metabolic processes can be quite slow (with characteristic times of the order of thousands of seconds), a single extracellular space has a very small volume and the time it takes to fill or empty this volume is very short (of the order of 10--100 $\mu$s). Thus the characteristic times span eight orders of magnitude, and the system is quite stiff. This means that any adopted integration method must be very stable, and that it should be able to overlook very short transients that do not mean much from a biophysical point of view, and it should also accept long time steps without crashing or producing unphysical results. 

All this proves quite challenging, as it poses conflicting requests on the integration algorithms. 
The usual considerations on stability \cite{dahl,butcher,NR} suggest that implicit methods should perform better than explicit methods, and this also means that a robust algorithm must sacrifice speed, as implicit methods require several function evaluations. 

Multistep methods \cite{butcher,NR,odeh}  may appear to be a smart choice, because they can incorporate information from previous simulation steps and thus reduce the number of required function evaluations: however numerical tests performed in anticipation of the present work have shown that explicit Adams-Bashforth multistep methods are highly unstable in this context, and that the fourth-order Adams-Moulton implicit multistep method does not fare much better as it seems to be very sensitive to the quality of the algorithm initialization steps (that must necessarily be performed with another integration algorithm), and unfortunately the discrete events lead to a constant flow of equation initialization processes. 

Finally we have decided to include the following methods in the tests summarized in the next section: 
\begin{itemize}
\item{the default integrator of {\it Mathematica} \cite{Mathbook}, as a benchmark algorithm;}
\item{the implicit Euler method;}
\item{the implicit trapezoidal method (which corresponds to the second-order Adams-Moulton algorithm).}
\end{itemize}

Although it cannot be included as it is in a simulation program, we have decided to choose the default setting of the {\it Mathematica} instruction {\tt NDSolve} as a benchmark since it incorporates a well-tuned mixture of standard algorithms not included in our very short list \cite{Mathbook}, and thus provides a qualitative guide to possible better choices. Here simplicity is an important bonus, since the integration algorithm must be incorporated in a larger structure and the biophysical model may change frequently, as new biochemical paths are incorporated: however we have not included the simple explicit Euler method in the list, because it is well-known to fail badly in the case of stiff equations, unless the step size is exceedingly small (and yet this method is widely used in many similar contexts, see, e.g, \cite{hb}). The many important Runge-Kutta (RK) methods are also missing, because the stable implicit RK methods are not easily adopted for inclusion in a simulation program like VBL (or other similar numerical stepping schemes), and also because the sophisticated integrator of {\it Mathematica} \cite{Mathbook} actually includes a choice of RK methods -- both explicit and implicit -- and its performance may suggest corrections to our selection of algorithms. 

Since stability -- and not precision -- is the main feature that an integrator must have in this context, it would be important to set the general problem in a form such that the standard theorems on stability can be applied \cite{butcher}. Unfortunately it is impossible to specify exactly the operating conditions of the simulator, especially because the number of equations is variable -- as cells proliferate -- and because the theoretical model must be open to changes as our understanding of the biophysical and biochemical processes evolves. However there are a couple of general considerations that can help: the first is that the systems of equations must be autonomous, time cannot appear explicitly. The second is that the equations describe the evolution of masses and concentrations, i.e. quantities that must be non-negative, and this means that the generic structure of the equations for non-negative quantities $x_k$ must be 
\begin{equation}
\label{generic}
\frac{dx_k}{dt} = A(x_1, \ldots, x_N; t) - B(x_1, \ldots, x_N; t)x_k
\end{equation}
where $A$ and $B$ are non-negative functions that express respectively production and consumption/destruction of the quantity $x_k$. The consumption/destruction term must be proportional to $x_k$ to ensure the continued non-negativity of the solution (if there were not such a proportionality, then the derivative could be nonvanishing and negative for  values of $x_k$ close to zero, and thus lead to negative $x_k$). Indeed the proportionality is often related to Michaelis-Menten terms with a generic structure $x_k/(K_m+x_k)$. A differential system such as (\ref{generic}) can be locally approximated by the simpler equations 
\begin{equation}
\frac{dx_k}{dt} = A_0 - B_0 x_k
\end{equation}
which shows that A-stable integration methods should normally be sufficient in this complex context 
\cite{butcher}.

\section{Numerical results on the test model of section \ref{bio}}
\label{res}

We have chosen parameter values derived from experiment or from fits of experimental data and they are listed in table 1; all parameters are further explained and referenced in \cite{VBL1,VBL2}, except $D$ \cite{VLK} and $\Delta$. The approximate value of $\Delta$ is derived from the estimate that the extracellular matrix amounts to about 20\% of the total volume of human tissue \cite{jin} and assuming a free space fraction of about 50\% \cite{jain} (we note that this may actually be an overestimate): then we obtain $\Delta \approx 0.5 \left( 0.25 V_{in}\right) /S \approx 0.2 \mu$m.   

Using the values in the table we can find the equilibrium values that correspond to ${{dm_{in} }}/{{dt}}=0$ and ${{dm_{extra} }}/{{dt}}=0$, i.e., $\left[ G_{in} \right] \approx 0.043$ kg m$^{-3}$  and $\left[ G_{extra} \right] \approx 0.90$ kg m$^{-3}$: this means that because of conversion into G6P, the glucose level in this isolated cell is considerably lower than in the surrounding environment. 

\begin{table}[htdp]
\label{table1}
\begin{center}
\begin{tabular}{|c|l|p{7cm}|}
\hline 
{\bf parameter} & {\bf value} &  \\
\hline 
$r$ & $5\cdot 10^{-6}$ m & cell radius \\
\hline 
$S=4\pi r^2$ & $3\cdot 10^{-10}$ m & surface area \\
\hline 
$V_{in} = \frac{4\pi}{3}r^3$ & $5 \cdot 10^{-16}$ m$^2$ & cell volume \\
\hline 
$\Delta$ & $0.2 \cdot 10^{-6}$ m & thickness of extracellular space around cell \\
\hline 
$V_{extra} = S \Delta$ & $6 \cdot 10^{-17}$ m$^3$ & extracellular volume \\
\hline $v_{max,1} \approx 2\cdot 10^{-9} S$ & $6\cdot 10^{-19}$ kg s$^{-1}$ & maximum speed of GLUT-mediated glucose transport \\
\hline
$v_{max,2}$ & $1.2 \cdot 10^{-19}$ kg s$^{-1}$ & maximum rate of glucokinase activity \\
\hline 
$v_{max,22}$ & $1.2 \cdot 10^{-18}$ kg s$^{-1}$ & maximum rate of hexokinase activity \\
\hline 
$K_1$ & $0.27024$ kg m$^{-3}$ & Michaelis-Menten constant of glucose transport \\
\hline 
$K_2$ & $1.8$ kg m$^{-3}$ & Michaelis-Menten constant for glucokinase \\
\hline 
$K_{22}$ & $1.8\cdot 10^{-2}$ kg m$^{-3}$ & Michaelis-Menten constant for hexokinase \\
\hline 
$K_a$ & $5.4\cdot 10^{-2}$ kg m$^{-3}$ & constant for the homeostatic loop controlling glucokinase and hexokinase activity\\
\hline 
$D$ & $7\cdot 10^{-10}$ m$^2$ s$^{-1}$ & diffusion constant of glucose in water \\
\hline 
$\left[ G_{env} \right]$ & $0.9$ kg m$^{-3}$ & glucose concentration in  standard nourishing solutions\\
\hline 
\end{tabular} 
\caption{Parameters used in the numerical tests with one isolated cell. Parameters are explained and referenced in \cite{VBL1,VBL2}, except $D$ \cite{VLK} and $\Delta$ (see discussion in the main text).}
\end{center}
\end{table}

As explained above, the system of equations (\ref{mass1}) and (\ref{mass2}) is very stiff, and this can be easily visualized setting an off-equilibrium initial glucose concentration inside the cell. Some of the glucose is quickly converted into G6P, but part of it either seeps out or enters the cell, and changes the concentration in the extracellular space. Since the extracellular volume is very small it reacts very quickly to this sudden change, and exchanges glucose with the environment, exhibiting a sharp transient. The approximate duration of this transient can be estimated expanding equation (\ref{mass2}) for very low values of the masses: since $V_{in} \gg V_{extra}$, 
\begin{equation}
\label{mass2approx}
 \frac{{dm_{extra} }}{{dt}} \simeq  - \left( \frac{{v_{\max ,1}  }}{{V_{extra} K_1 } } +\frac{{DS}}{V_{extra} \Delta } \right) m_{extra} 
 + \frac{{DS}}{\Delta }\left[ {G_{env} } \right] ,
\end{equation}
with the parameters of table 1, this gives $m_{extra}(t) \sim A + B\exp\left(-t/\tau \right)$, with $\tau \approx 60 \mu$s. It is also important to note that this very short time is entirely due to the diffusion term in equation (\ref{mass2approx}): without the diffusion term the characteristic time $\tau$ would grow to more than 27 s.

The nearly exponential transient is clearly visible in a plot of the mass flow into the extracellular space
\begin{equation}
\left.  \frac{{dm_{extra} }}{{dt}}\right|_{inflow} =\frac{{DS}}{\Delta }\left( {\left[ {G_{env} } \right] - \frac{{m_{extra} }}{{V_{extra} }}} \right) 
 \end{equation}
which is shown in figure \ref{fig1} for the initial conditions $m_{in}=0.1\left[ {G_{env} } \right] V_{in} $ and $m_{extra}=0.1\left[ {G_{env} } \right] V_{extra}$. Because of this fast transient, many standard algorithms fail when integration is carried out on long time intervals with comparatively long time steps: rather surprisingly, {\it Mathematica} itself fails badly (at least with the standard settings for {\tt NDSolve}) even though it includes a stiffness detection procedure \cite{Mathbook}. This failure becomes slowly apparent as the integration is carried out on longer and longer intervals; figure {\ref{fig2}) shows the concentration $\left[ G_{in} \right]$ vs. time for a 1000 seconds interval and initial conditions $m_{in}=0.9\left[ {G_{env} } \right] V_{in} $ and $m_{extra}=0.1\left[ {G_{env} } \right] V_{extra}$: instead of a smooth decay towards the equilibrium value the {\it Mathematica} solution shows some unexpected undulations. The companion plot for the concentration in the extracellular space $\left[ G_{extra} \right]$ (not shown) displays a large, unphysical peak close to the origin (this peak is about 100 times larger than the environmental concentration). A similar integration performed on a longer interval (10000 s) returns an even worse result, as it produces negative and unphysical values for the concentration  $\left[ G_{in} \right]$ (see figure \ref{fig3}). 

\begin{figure}
\begin{center}
\includegraphics[width=14cm]{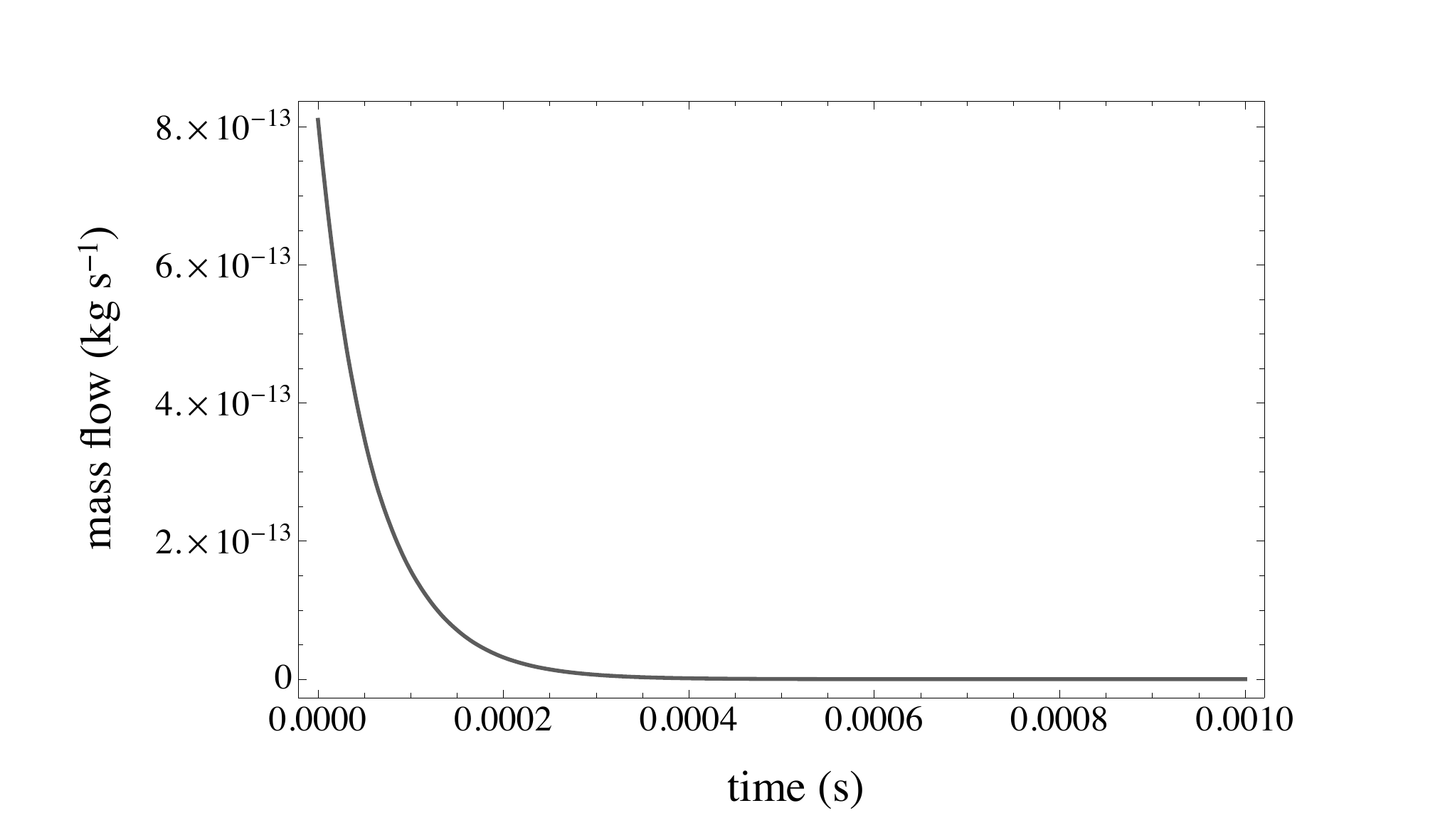}
\caption{\label{fig1}  Mass flow into the extracellular space (mass flow (kg s$^{-1}$) vs. time (s)) obtained from the numerical integration of equations (\ref{mass1}) and (\ref{mass2}) with {\it Mathematica}. A direct fit of this nearly exponential curve yields a decay time $\tau \approx 62 \mu$s, very close to the estimate in the main text.}
\end{center}
\end{figure}

\begin{figure}
\begin{center}
\includegraphics[width=14cm]{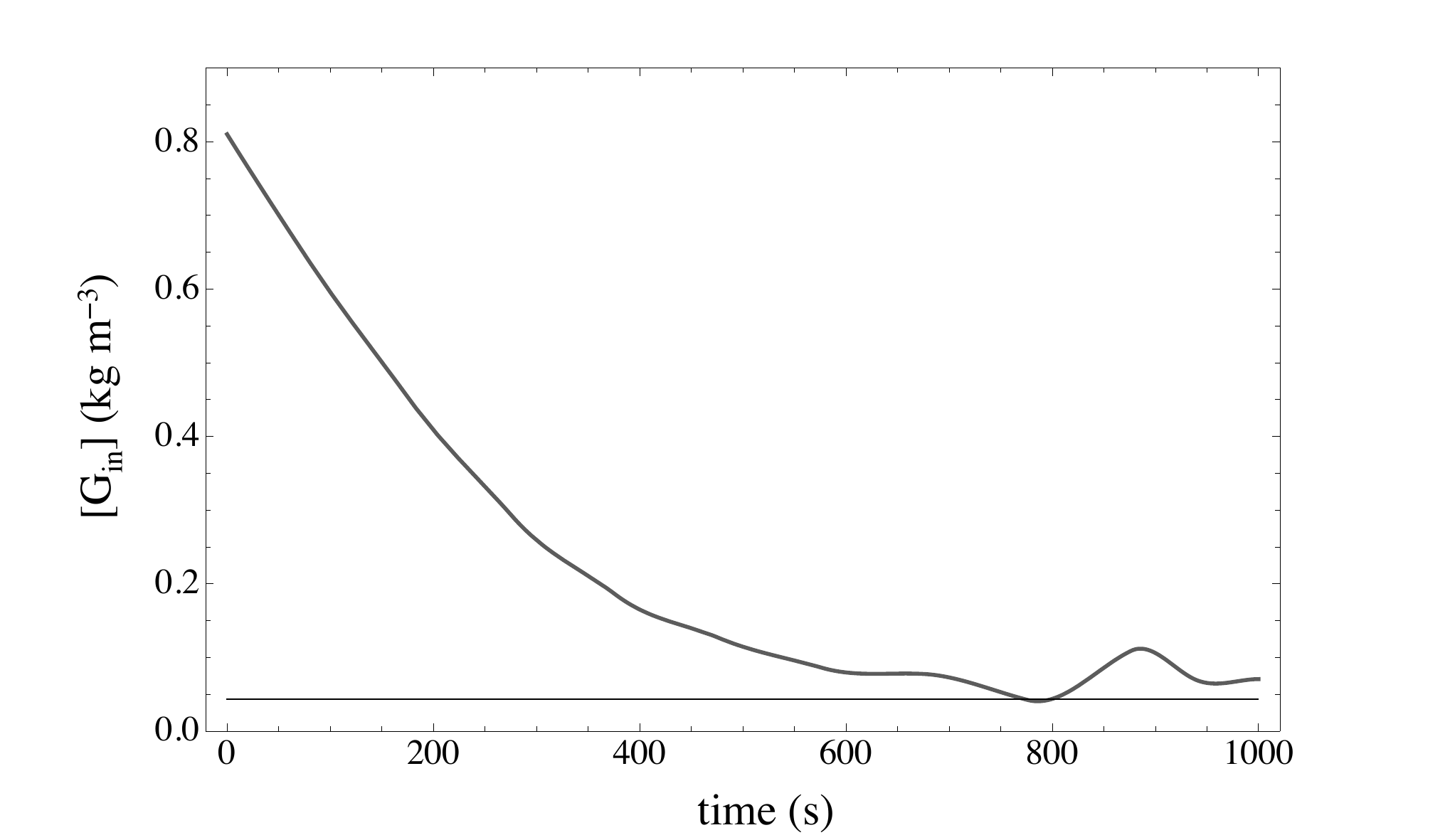}
\caption{\label{fig2}  Plot of the concentration $\left[ G_{in} \right]$ (kg m$^{-3}$) vs. time (s) obtained from the numerical integration of equations (\ref{mass1}) and (\ref{mass2}) with {\it Mathematica} for a 1000 seconds long time interval. The gray horizontal line shows the equilibrium level. The internal stepping procedure (with the default values) fails and produces unexpected undulations.}
\end{center}
\end{figure}

\begin{figure}
\begin{center}
\includegraphics[width=14cm]{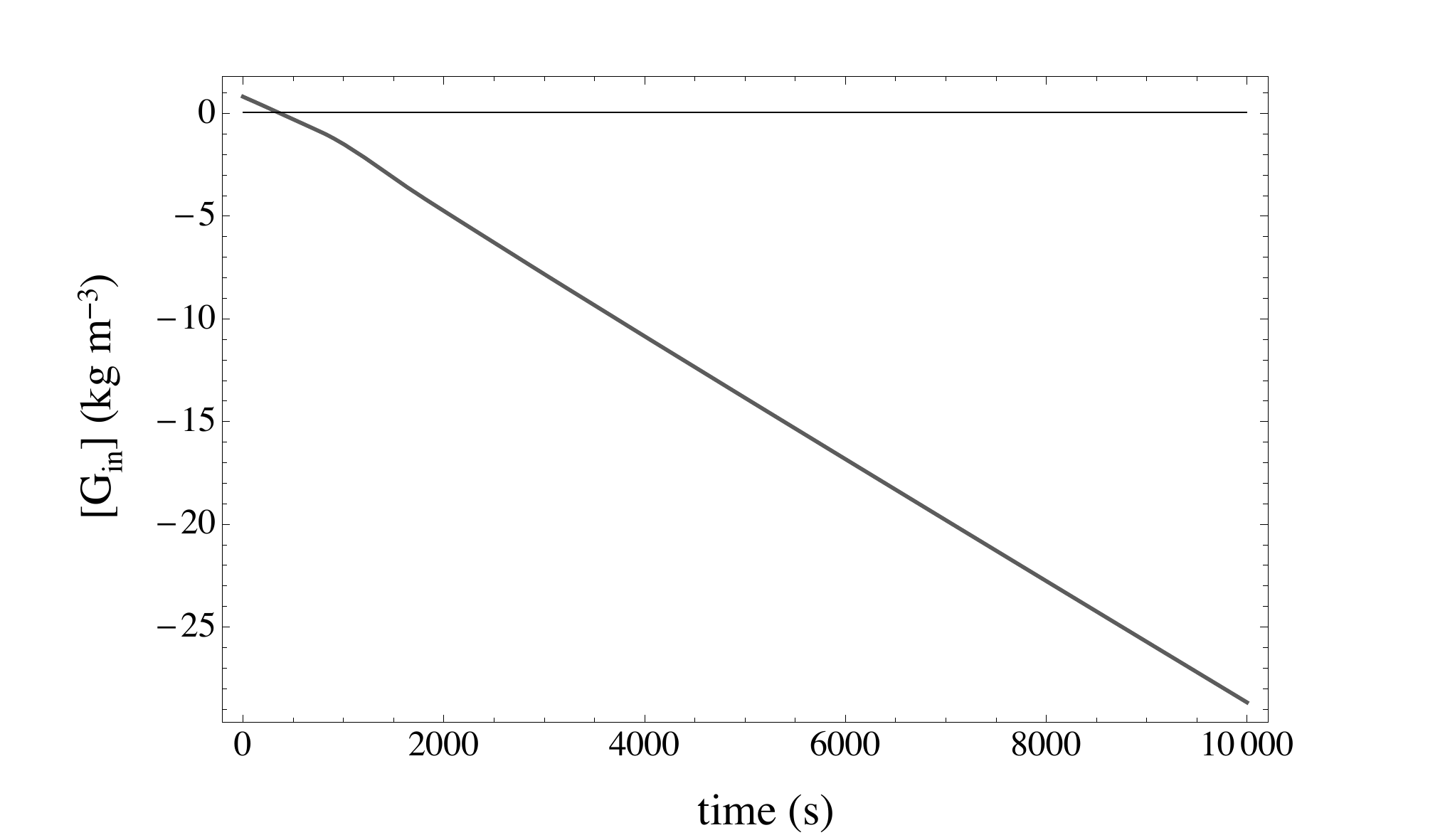}
\caption{\label{fig3}  Plot of the concentration $\left[ G_{in} \right]$ (kg m$^{-3}$) vs. time (s) obtained from the numerical integration of equations (\ref{mass1}) and (\ref{mass2}) with {\it Mathematica} for a 10000 seconds long time interval. The gray horizontal line shows the equilibrium level. The internal stepping procedure (with the default values) fails badly and leads to large, negative (unphysical) values of the concentration $\left[ G_{in} \right]$.}
\end{center}
\end{figure}

\begin{figure}
\begin{center}
\includegraphics[width=14cm]{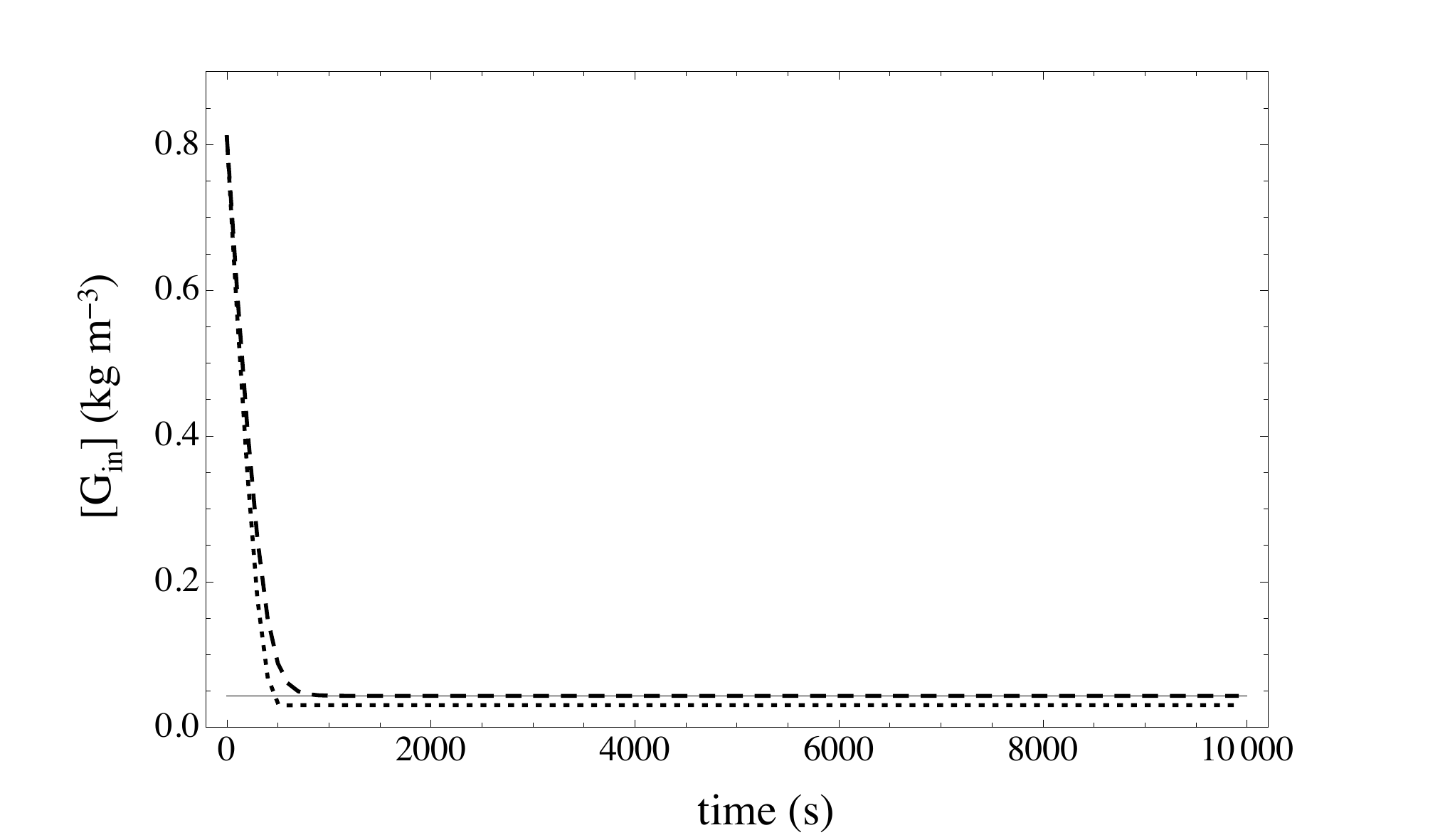}
\caption{\label{fig4}  Plot of the concentration $\left[ G_{in} \right]$ (kg m$^{-3}$) vs. time (s) obtained from the numerical integration of equations (\ref{mass1}) and (\ref{mass2}) with the implicit Euler method (dashed curve) and with the implicit trapezoidal method (dotted curve) for a 10000 seconds long time interval; in each case the step size is 100 s. The gray horizontal line shows the equilibrium level: the implicit Euler method reaches this equilibrium value without unwanted oscillations, while the implicit trapezoidal method shoots below the line and afterwards climbs very slowly back to the equilibrium value. We find that this climb depends on the actual stepsize, and it is faster for shorter time steps. Neither algorithm produces unphysical values (negative concentrations).}
\end{center}
\end{figure}

A careful analysis of the curve in figure \ref{fig1} reveals that the fast transient is very close to an exponential, and therefore we expect the standard results on A-stability of integration methods \cite{butcher} to be applicable to this and to similar nonlinear differential systems. In particular, we know that both the implicit Euler and the implicit trapezoidal method are unconditionally A-stable. Indeed the tests performed show that both methods converge even when the internal machinery of {\it Mathematica} -- which uses a full array of sophisticated methods -- is foiled by this stiff system. And although it is known that the accuracy of the implicit trapezoidal method (IT) is higher than that of the implicit Euler algorithm (IE) \cite{butcher}, in these tests IE fares better, since the IT method shoots below the equilibrium value of $\left[ G_{in} \right]$ before swinging -- very slowly back (see figure \ref{fig4}),  and displays unwanted oscillations of $\left[ G_{extra} \right]$ that very gradually  die out. The IT method also alternates steps above and below the equilibrium value of $\left[ G_{extra} \right]$, and these oscillations depend on the step size and resemble those observed in a closely related method, the Crank-Nicolson method for partial differential equations \cite{gersh,bri}.

\section{Diffusion in cell clusters}
\label{diff}

In the previous section we have considered just one cell, but as explained above, we wish to simulate large clusters of cells, and this means that we have to deal with large systems of coupled nonlinear differential equations. We begin with a very small cluster of just two cells, where glucose dynamics is described by the following equations for the masses
\begin{eqnarray}
\nonumber
 \frac{{dm_{in}^{(A)} }}{{dt}} &=& \frac{{v^{(A)} _{\max ,1} m_{extra}^{(A)} }}{{V_{extra}^{(A)} K_1  + m_{extra}^{(A)} }} - \frac{{v^{(A)} _{\max ,1} m_{in}^{(A)} }}{{V_{in}^{(A)} K_1  + m_{in}^{(A)} }} + \\
\nonumber
 &&
 \frac{{v^{(AB)} _{\max ,1} m_{extra}^{(AB)} }}{{V^{(AB)} _{extra} K_1  + m_{extra}^{(AB)} }} - \frac{{v^{(AB)} _{\max ,1} m_{in}^{(A)} }}{{V_{in}^{(A)} K_1  + m_{in}^{(A)} }} \\ 
\nonumber
  &&
  - \frac{{v_{\max ,2} \left( {m_{in}^{(A)} } \right)^2 }}{{\left( {V_{in}^{(A)} K_2  + m_{in}^{(A)} } \right)\left( {V_{in}^{(A)} K_a  + m_{in}^{(A)} } \right)}} \\
\label{mass2-1}
&&
- \frac{{v_{\max ,22} \left( {m_{in}^{(A)} } \right)^2 }}{{\left( {V_{in}^{(A)} K_{22}  + m_{in}^{(A)} } \right)\left( {V_{in}^{(A)} K_a  + m_{in}^{(A)} } \right)}} \\ 
\nonumber
 \frac{{dm_{in}^{(B)} }}{{dt}} &=& \frac{{v^{(B)} _{\max ,1} m_{extra}^{(B)} }}{{V^{(B)} _{extra} K_1  + m_{extra}^{(B)} }} - \frac{{v^{(B)} _{\max ,1} m_{in}^{(B)} }}{{V_{in}^{(B)} K_1  + m_{in}^{(B)} }} \\
 \nonumber
 &&
 + \frac{{v^{(AB)} _{\max ,1} m_{extra}^{(AB)} }}{{V_{in}^{(AB)} K_1  + m_{extra}^{(AB)} }} - \frac{{v^{(AB)} _{\max ,1} m_{in}^{(B)} }}{{V_{in}^{(B)} K_1  + m_{in}^{(B)} }} \\ 
 \nonumber
 &&
  - \frac{{v_{\max ,2} \left( {m_{in}^{(B)} } \right)^2 }}{{\left( {V_{in}^{(B)} K_2  + m_{in}^{(B)} } \right)\left( {V_{in}^{(B)} K_a  + m_{in}^{(B)} } \right)}} \\
\label{mass2-2}
&&
- \frac{{v_{\max ,22} \left( {m_{in}^{(B)} } \right)^2 }}{{\left( {V_{in}^{(B)} K_{22}  + m_{in}^{(B)} } \right)\left( {V_{in}^{(B)} K_a  + m_{in}^{(B)} } \right)}} \\ 
\nonumber
 \frac{{dm_{extra}^{(A)} }}{{dt}} &=&  - \frac{{v^{(A)} _{\max ,1} m_{extra}^{(A)} }}{{V_{extra}^{(A)} K_1  + m_{extra}^{(A)} }} + \frac{{v^{(A)} _{\max ,1} m_{in}^{(A)} }}{{V_{in}^{(A)} K_1  + m_{in}^{(A)} }} \\
 \label{mass2-3}
&&
 + \frac{{DS^{(A)} }}{\Delta }\left( {\left[ {G_{env} } \right] - \frac{{m_{extra}^{(A)} }}{{V_{extra}^{(A)} }}} \right) \\  
\nonumber
 \frac{{dm_{extra}^{(B)} }}{{dt}} &=&  - \frac{{v^{(B)} _{\max ,1} m_{extra}^{(B)} }}{{V^{(B)} _{extra} K_1  + m_{extra}^{(B)} }} + \frac{{v^{(B)} _{\max ,1} m_{in}^{(B)} }}{{V_{in}^{(B)} K_1  + m_{in}^{(B)} }} \\
\label{mass2-4}
&&
+ \frac{{DS^{(B)} }}{\Delta }\left( {\left[ {G_{env} } \right] - \frac{{m_{extra}^{(B)} }}{{V_{extra}^{(B)} }}} \right) \\ 
\nonumber
 \frac{{dm_{extra}^{(AB)} }}{{dt}} &=&  - 2\frac{{v^{(AB)} _{\max ,1} m_{extra}^{(AB)} }}{{V^{(AB)} _{extra} K_1  + m_{extra}^{(AB)} }} + \frac{{v^{(AB)} _{\max ,1} m_{in}^{(A)} }}{{V_{in}^{(A)} K_1  + m_{in}^{(A)} }} + \frac{{v^{(AB)} _{\max ,1} m_{in}^{(B)} }}{{V_{in}^{(B)} K_1  + m_{in}^{(B)} }} \\
\label{mass2-5}
 &&
 + \frac{{DS^{(L)} }}{{\Delta _L }}\left( {\left[ {G_{env} } \right] - \frac{{m_{extra}^{(AB)} }}{{V_{extra}^{(AB)} }}} \right)
\end{eqnarray}
where the superscripts $(A)$ and $(B)$ denote the two cells, and the parameters in the numerical tests are a straightforward extension of those used for a single cell and are listed in table 2. 

\begin{table}[htdp]
\label{table2}
\begin{center}
\begin{tabular}{|c|l|p{7cm}|}
\hline 
{\bf parameter} & {\bf value} &  \\
\hline 
$r^{(A)} = r^{(B)} = r$ & $5\cdot 10^{-6}$ m & cell radius \\
\hline 
$\Delta$ & $0.2 \cdot 10^{-6}$ m & thickness of extracellular space around cell \\
\hline 
$\Delta_L = r$ & $5 \cdot 10^{-6}$ m & effective distance for calculation of flow into and out of the extracellular space between cells\\
\hline 
$S=4\pi r^2$ & $3\cdot 10^{-10}$ m & total surface area \\
\hline 
$S^{(A)} = S^{(B)} = S/2$ & $1.5 \cdot 10^{-10}$ m & surface area of each hemisphere \\
\hline 
$S^{(AB)} = \pi r^2$ & $7.9 \cdot 10^{-11}$ m & contact area between cells \\
\hline 
$S^{(L)}$ & $1.5 \cdot 10^{-10}$ m & surface area of each hemisphere \\
\hline 
$V_{in}^{(A)} = V_{in}^{(B)} = \frac{2\pi}{3}r^3$ & $2.5 \cdot 10^{-16}$ m$^2$ & cell volume \\
\hline 
$V_{extra}^{(A)} = V_{extra}^{(B)} = 2\pi r^2 \Delta$ & $3 \cdot 10^{-17}$ m$^2$ & volume of each hemispherical extracellular space\\
\hline 
$V_{extra}^{(AB)} = \pi r^2 \Delta$ & $1.6 \cdot 10^{-17}$ m$^2$ & volume of extracellular space between cells \\
\hline 
$v_{max,1}^{(A)} =v_{max,1}^{(B)}$ & $3\cdot 10^{-19}$ kg s$^{-1}$ & maximum speed of GLUT-mediated glucose transport between each cell and surrounding hemispherical extracellular space \\
\hline 
$v_{max,1}^{(AB)}$ & $1.6\cdot 10^{-19}$ kg s$^{-1}$ & maximum speed of GLUT-mediated glucose transport between each cell and the extracellular space between cells\\
\hline
$v_{max,2}$ & $1.2 \cdot 10^{-19}$ kg s$^{-1}$ & maximum rate of glucokinase activity \\
\hline 
$v_{max,22}$ & $1.2 \cdot 10^{-18}$ kg s$^{-1}$ & maximum rate of hexokinase activity \\
\hline 
$K_1$ & $0.27024$ kg m$^{-3}$ & Michaelis-Menten constant of glucose transport \\
\hline 
$K_2$ & $1.8$ kg m$^{-3}$ & Michaelis-Menten constant for glucokinase \\
\hline 
$K_{22}$ & $1.8\cdot 10^{-2}$ kg m$^{-3}$ & Michaelis-Menten constant for hexokinase \\
\hline 
$K_a$ & $5.4\cdot 10^{-2}$ kg m$^{-3}$ & constant for the homeostatic loop controlling glucokinase and hexokinase activity\\
\hline 
$D$ & $7\cdot 10^{-10}$ m$^2$ s$^{-1}$ & diffusion constant of glucose in water \\
\hline 
$\left[ G_{env} \right]$ & $0.9$ kg m$^{-3}$ & glucose concentration in  standard nourishing solutions\\
\hline 
\end{tabular} 
\caption{Parameters used in the numerical tests with two cells.}
\end{center}
\end{table}

The geometric structure of the two cells is depicted in figure \ref{fig5}: this structure approximates the shape of two cells just after mitosis, and we distinguish six different regions
\begin{itemize}
\item{the two hemispherical cells;}
\item{two hemispherical extracellular spaces that are the interface between the cells and the environment;}
\item{the cylindrical extracellular space between the cells; this cylindrical space is in contact with the environment only through a narrow belt; }
\item{the (fixed) environment.}
\end{itemize}

This rough division of space contains the main elements of simulations where diffusion plays an important role, and later in this section we shall see how to expand it. 

\begin{figure}
\begin{center}
\includegraphics[width=14cm]{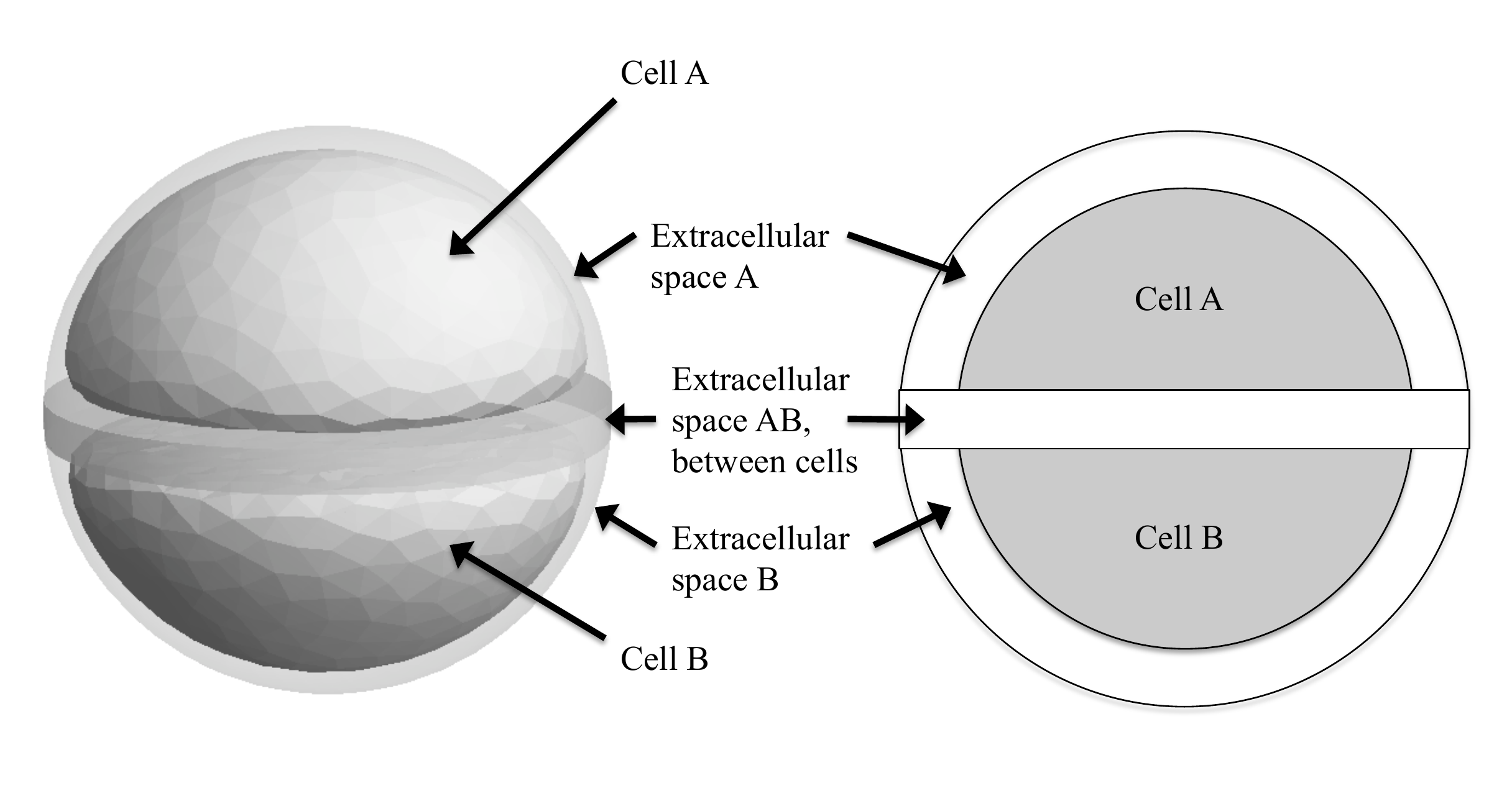}
\caption{\label{fig5}  Schematic view of the geometric structure of the pair of daughter cells described in the main text, where it is assumed that just after mitosis the cells are (nearly) hemispherical and are surrounded by two hemispherical extracellular spaces -- the interface with the environment -- and are separated by one cylindrical extracellular space. Left panel: perspective view; right panel: cross section. For clarity the thickness of the extracellular spaces is exaggerated; the interface between extracellular spaces is actually negligible and is not counted in the equations (\ref{mass2-1}) to (\ref{mass2-5}).}
\end{center}
\end{figure}

We have carried out extensive numerical tests similar to those in the previous section, and essentially they confirm what we found with a single cell: 
\begin{itemize}
\item{the {\it Mathematica} solution obtained with the default settings fails for long integration intervals;}
\item{the implicit Euler method and the implicit trapezoidal method both converge to the equilibrium value, but the implicit trapezoidal method displays strong and very undesirable oscillations.}
\end{itemize}

This small system with just two cells can be extended further, with the addition of more cells and their extracellular spaces: this is equivalent to the discrete volume version of a diffusion problem (see, e.g., \cite{saad} for a review of volume discretization for the solution of diffusion problems), i.e., we must deal with a large system of equations (here we use concentrations $\rho$ rather than the masses because the ensuing analysis is slightly easier to follow): 
\begin{equation}
\label{diffsys}
V_a \frac{{d\rho _a }}{{dt}} = F_a\left( {\rho_a;\rho_a } \right) + D\sum\limits_{\left\langle b \right\rangle } \left( \rho_a  - \rho_b  \right)g_{ab} 
\end{equation}
where indexes denote both cells and extracellular spaces, $g_{ab}$ is term related to geometry (it is equivalent to the $S/\Delta$ ratio that we met previously), the $F_a$'s describe facilitated transport (for both cells and extracellular spaces) and metabolic processes (cells only), and the diffusion term is actually included only in the case of extracellular spaces. The subscript $\left\langle b \right\rangle$ indicates that the sum is performed only on the set of cells $b$ that are adjacent to cell $a$: since in a simulator like VBL the cells are in arbitrary positions in space \cite{VBL0}, and not on the usual cubic lattice, the number of neighbors is random (in a 3D configuration it fluctuates about an average of 12).

We have already remarked in the previous section that the $F_a$'s are slowly varying functions and that the stiffness of the differential system and the algorithmic stability problems are almost entirely due to the diffusion terms: for this reason we concentrate on a reduced version of the differential system (\ref{diffsys}), i.e., 
\begin{equation}
\label{diffsysred}
 \frac{{d\rho _a }}{{dt}} \approx \frac{D}{V_a}\sum\limits_{\left\langle b \right\rangle } \left( \rho_a  - \rho_b  \right)g_{ab} \end{equation}
It is well-known that differential systems like (\ref{diffsysred}), obtained from the discretization of diffusion problems on regular lattices, can be integrated by implicit methods like the Backward Differentiation Formulas (BDF) or the Crank-Nicolson algorithm (CN) \cite{NR,gersh}, and that in those cases both BDF and CN are unconditionally stable \cite{gersh}. However in simulation programs like VBL, the equations are not discretized on a lattice, and may have a variable number of diffusion terms, as the sum runs over a random number of neighbors: are these implicit integration methods still stable in this more general setting? Fortunately the answer is yes, because it is possible to develop a variation of the standard Von Neumann stability analysis. The argument for BDF goes as follows: we start from the discrete-time version of (\ref{diffsysred}) that corresponds to the BDF iterations
\begin{equation}
\label{diffsysBDF}
\rho_a^{n+1}  = \rho_a^{n} +  \Delta t \frac{D}{V_a}  \sum\limits_{\left\langle b \right\rangle } \left( \rho _b^{n+1}  - \rho _a^{n+1} \right)g_{ab} 
\end{equation}
where $\Delta t$ is the time step, $\rho_a^n = \rho_a(n\Delta t)$, and we take the test solution $\rho_a^n \simeq A^n \exp\left( i\mathbf{k}\cdot \mathbf{r}_a \right)$ that corresponds to the test solution used in the standard Von Neumann stability analysis, but without the assumption of a regular spatial lattice. Substituting the test solution in the iteration formulas (\ref{diffsysBDF}) we find
\begin{equation}
\label{ABDF}
\nonumber
A^{-1} =  1 + \frac{D\Delta t}{V_a } \sum\limits_{\left\langle b \right\rangle } g_{ab} \left( 1 - \cos \phi_{ab} \right)  
- i \frac{D\Delta t}{V_a } \sum\limits_{\left\langle b \right\rangle } g_{ab} \sin \phi_{ab}
\end{equation}
where $\phi_{ab} = \bf{k}\cdot\left( \bf{r}_b  - \bf{r}_a  \right)$, and since 
$\sum\limits_{\left\langle b \right\rangle } {g_{ab} \left( {1 - \cos \phi_{ab}} \right)} \ge 0$ we see that $|A| \le 1$ and thus the BDF algorithm is unconditionally stable. 

The CN iteration formula is 
\begin{equation}
\label{diffsysCN}
\rho_a^{n+1}  = \rho_a^{n} +  \Delta t \frac{D}{2 V_a} \left[  \sum\limits_{\left\langle b \right\rangle } \left( \rho _b^{n+1}  - \rho _a^{n+1} \right)g_{ab} + \sum\limits_{\left\langle b \right\rangle } \left( \rho _b^{n}  - \rho _a^{n} \right)g_{ab} \right]
\end{equation}
and we can repeat the steps followed for the BDF algorithm and find 
\begin{equation}
\label{ACN}
A = \frac{{1 - \frac{{D\Delta t}}{{V_a }}\sum_{\left\langle b \right\rangle } {g_{ab} \left( {1 - \cos \phi_{ab}} \right)}  
+ i\frac{{D\Delta t}}{{V_a }}\sum_{\left\langle b \right\rangle } {g_{ab} \sin \phi_{ab}} }} {{1 + \frac{{D\Delta t}}{{V_a }}\sum_{\left\langle b \right\rangle } {g_{ab} \left( {1 - \cos \phi_{ab}} \right)}  
- i\frac{{D\Delta t}}{{V_a }}\sum_{\left\langle b \right\rangle } {g_{ab} \sin \phi_{ab}} }}
\end{equation}
and just as before we see that $|A| \le 1$ and that the CN algorithm is unconditionally stable. 

\begin{figure}
\begin{center}
\includegraphics[width=14cm]{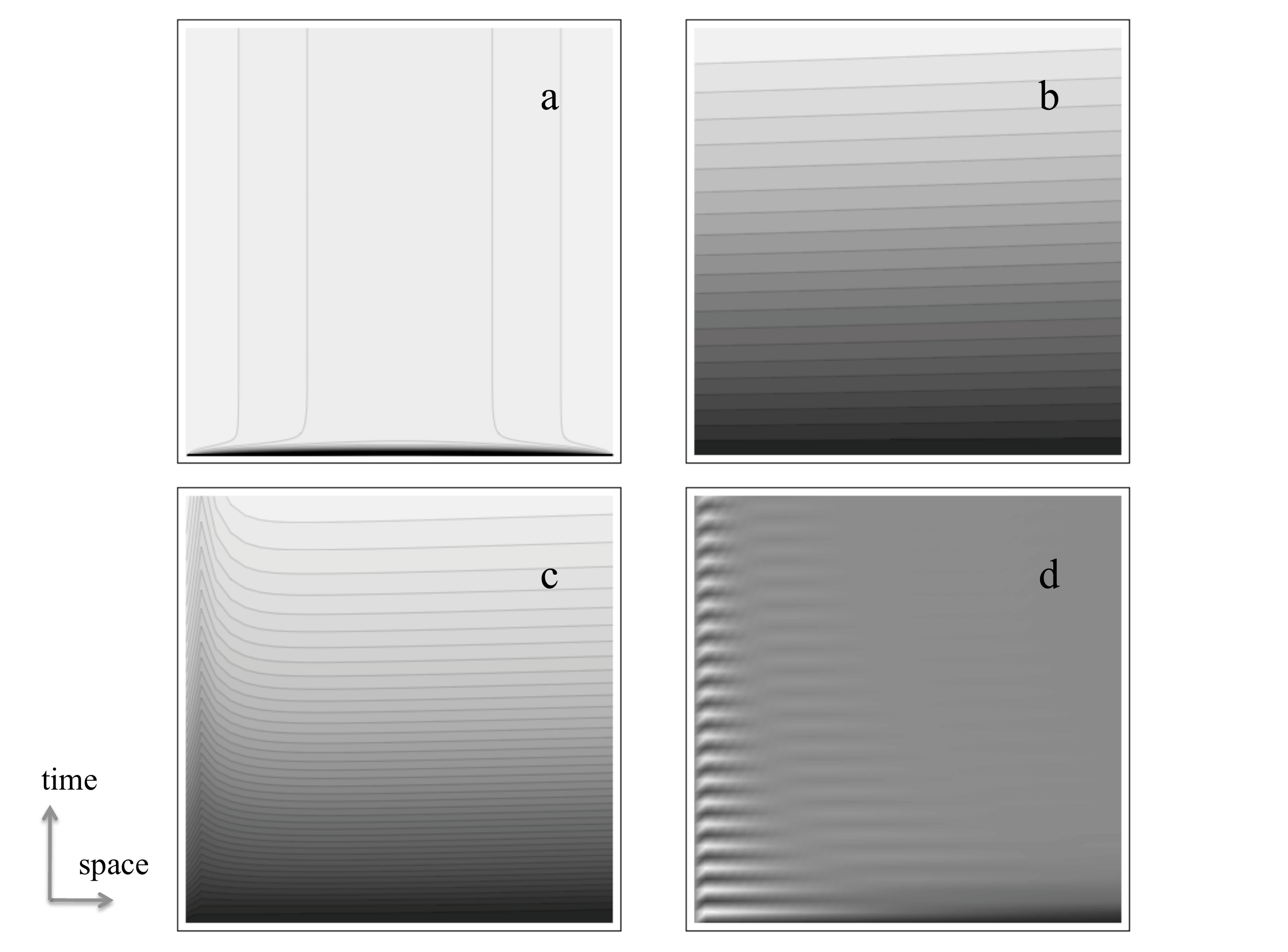}
\caption{\label{fig6}  Contour/density plots that show different aspects of the solutions of equations (\ref{string1}-\ref{string2}) for a string of 100 cells, that spans a spatial range $(-500 \mu\mathrm{m}, 500 \mu\mathrm{m})$. Space runs along the {\it x}-axis and time runs upwards along the {\it y}-axis with timesteps $\Delta t = 2$ s, and concentration values are also mapped on a gray scale (black correspond to the minimum value, white to maximum). The panels show: {\bf a)} the contour plot of the glucose concentration in the extracellular spaces obtained with the BDF method; {\bf b)} contour plot of the glucose concentrations inside cells, for cells in the reduced spatial range $(-500 \mu\mathrm{m}, -200 \mu\mathrm{m})$, for $t = 0$ s until $t = 60 $ s, obtained with the BDF method; {\bf c)} same as {\bf b}, but in this case the numerical solution has been calculated with the CN method: notice that the contour lines are appreciably distorted close to the boundary; {\bf d)} contour plot of the glucose concentrations in the extracellular spaces in the reduced spatial range $(-500 \mu\mathrm{m}, -200 \mu\mathrm{m})$, for $t = 0$ s until $t = 60 $ s, obtained with the CN method: notice the banded structure that is due to the oscillatory instability of the CN method. }
\end{center}
\end{figure}

\begin{figure}
\begin{center}
\includegraphics[width=14cm]{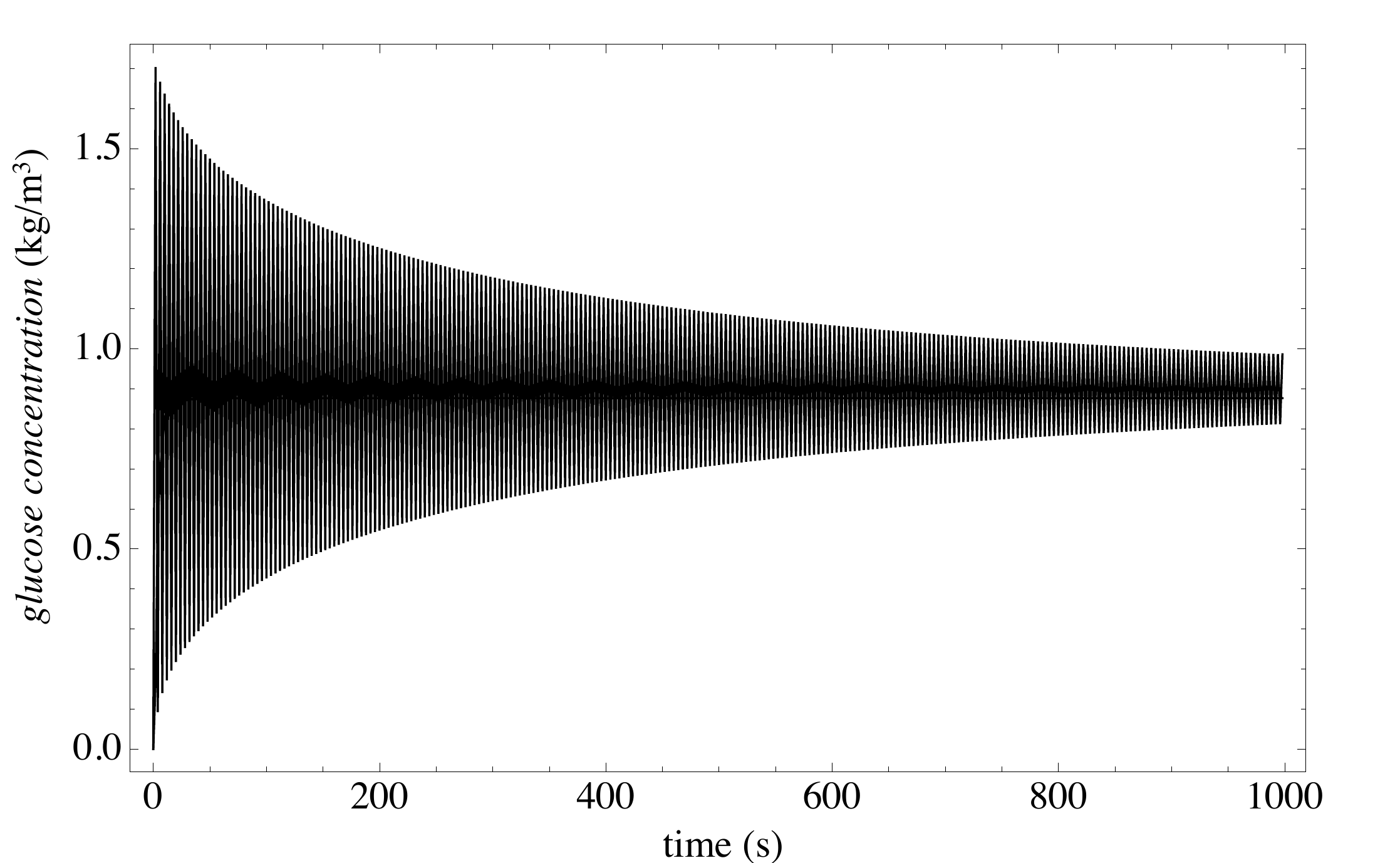}
\caption{\label{fig7}  Plot of the glucose concentration in the leftmost extracellular space in the string of cells described in the text, obtained numerically with the CN method for the same problem of figure \ref{fig6}. In this solution there is a marked high frequency spurious oscillation that depends on the time step and that very slowly fades away.}
\end{center}
\end{figure}

\begin{figure}
\begin{center}
\includegraphics[width=14cm]{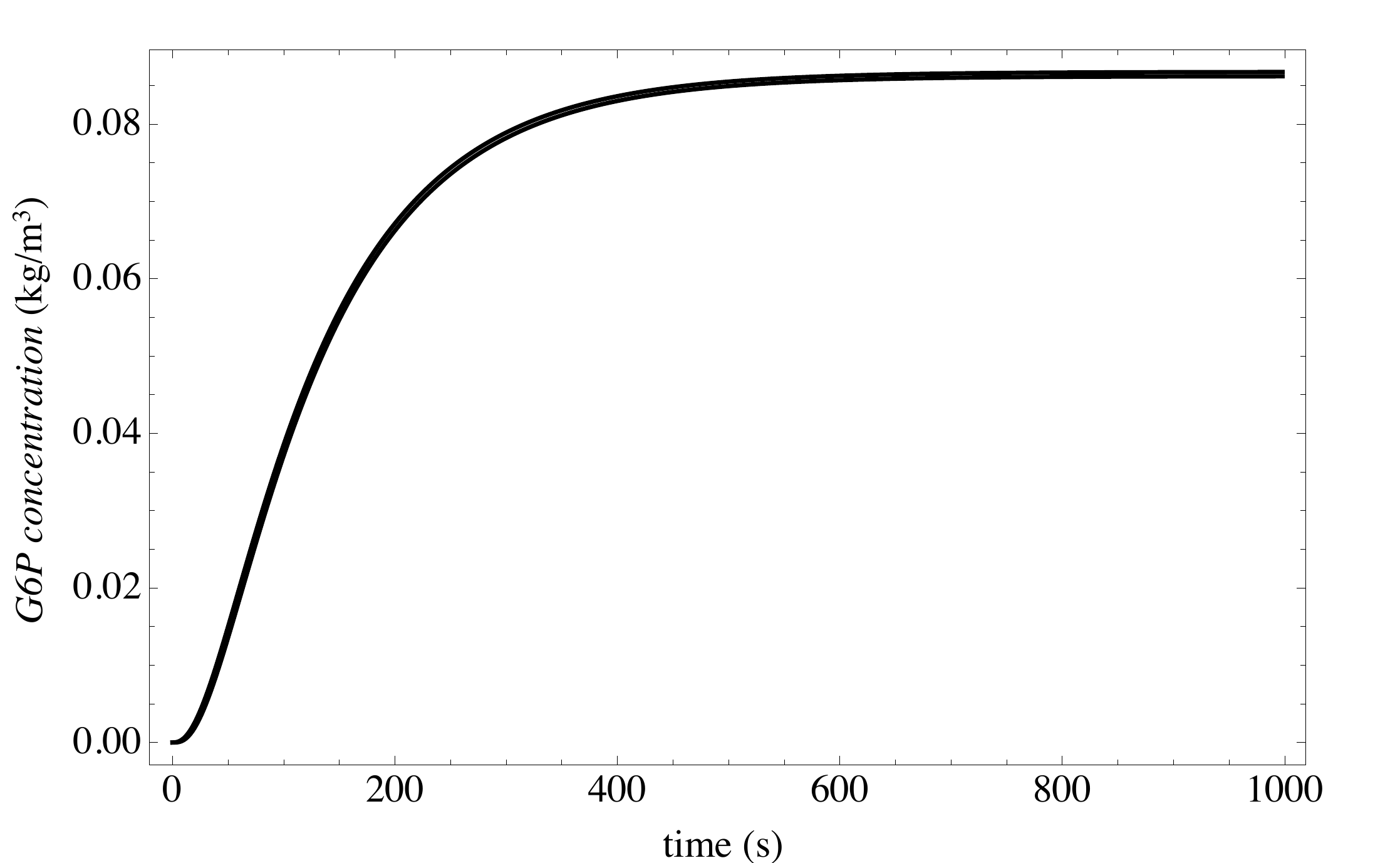}
\caption{\label{fig8}  Plot of G6P concentration at two positions (close to the boundary and in the middle of the string of cells), obtained numerically with the BDF method for the same problem of figure \ref{fig6}, using the differential equation for G6P in \cite{VBL1}, solved with the implicit Euler method; the concentration of G6P is almost independent of position along the string and the two curves are indistinguishable on this scale.}
\end{center}
\end{figure}

\begin{figure}
\begin{center}
\includegraphics[width=14cm]{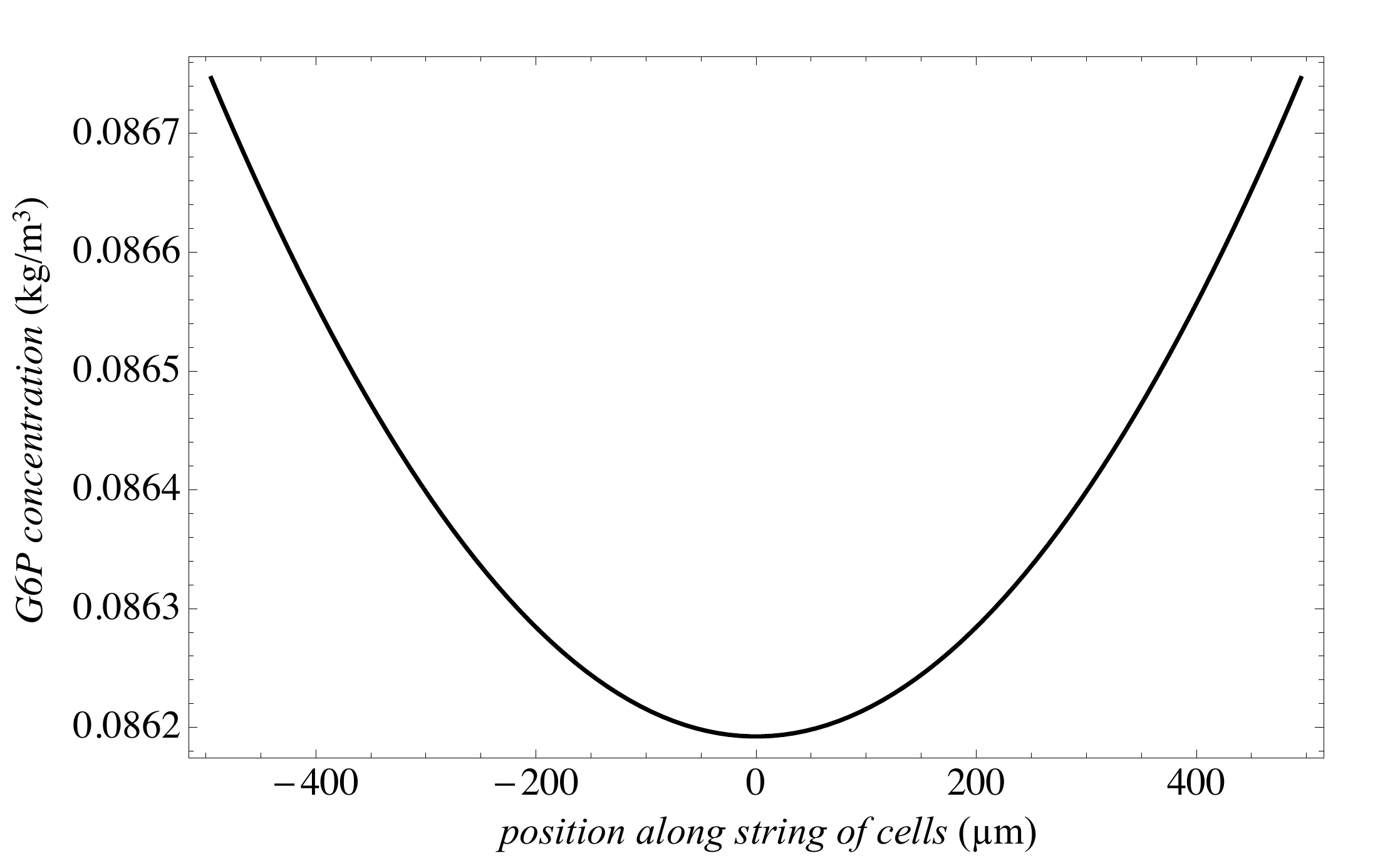}
\caption{\label{fig9}  Plot of G6P concentration along the string of cells at $t = $ 1000 s, obtained numerically with the BDF method for the same problem of figure \ref{fig6}, using the differential equation for G6P in \cite{VBL1}, solved with the implicit Euler method. The concentration is lowest in the center, and it mirrors a similar curve for glucose concentration. }
\end{center}
\end{figure}

To test the BDF and the CN algorithm in a realistic biophysical setting, we  have extended the two-cell model to a 1D string of cells, where we assume that each cell is surrounded by its own extracellular space, and that adjacent extracellular spaces can exchange material by diffusion. This means that the diffusion problem becomes slightly more complex (although the previous stability analysis still holds) and the equations for each cell/extracellular space are
\begin{eqnarray}
\label{string1}
 V_C \frac{{d\rho _C }}{{dt}} & = & M\left( {\rho _C } \right) + T\left( {\rho _c ,\rho _C } \right) \\ 
 \label{string2}
 V_c \frac{{d\rho _c }}{{dt}} & = & D\sum\limits_{\left\langle b \right\rangle } {\left( {\rho _b  - \rho _c } \right)g_{bc} }  - T\left( {\rho _c ,\rho _C } \right) 
\end{eqnarray}
where
\begin{eqnarray}
\nonumber
 M_G \left( {\rho_{G,C} } \right) & = & - \frac{{v_{\max ,2} \rho^2 _{G,C} }}{{\left( {K_2  + \rho_{G,C} } \right)\left( {K_a  + \rho_{G,C} } \right)}} \\
 &&- \frac{{v_{\max ,22} \rho^2 _{G,C} }}{{\left( {K_{22}  + \rho_{G,C} } \right)\left( {K_a  + \rho_{G,C} } \right)}} \\ 
 T_G \left( {\rho_{G,C} ,\rho_{G,c} } \right) & = & \frac{{v_{\max ,1} \rho_{G,c} }}{{K_1  + \rho_{G,c} }} - \frac{{v_{\max ,1} \rho_{G,C} }}{{K_1  + \rho_{G,C} }}
\end{eqnarray}
are the functions that describe the internal metabolic activity (conversion of glucose to G6P) and the transport processes into and out of cells and that we met earlier both in the one-cell and in the two-cell models. 
The uppercase subscripts denote cells and the lowercase subscripts denote extracellular spaces, and all the model parameters have already been listed in table 2. 
The space coordinate (cell positions) span the range $(x_{min}, x_{max})$, with the boundary conditions 
$u\left( { x_{min},t} \right) = u\left( {x_{max},t} \right) = $ constant: thus this is a model with both diffusion of glucose and local absorption and conversion into G6P. 

The BDF and the CN algorithm are equivalent to the implicit Euler and to the implicit trapezoidal method, respectively (they are actually the extensions of these algorithms to partial differential equations), and thus -- unsurprisingly -- they share with their counterparts both advantages and disadvantages. In particular in our tests on this particular diffusion problem we find that:
\begin{itemize}
\item{both BDF and CN eventually converge to the same equilibrium solution;}
\item{the BDF method requires a larger number of function evaluations to reach the required accuracy;}
\item{the CN method displays unwanted oscillations, similar to those found in the case of the implicit trapezoidal method.}
\end{itemize}
Figure \ref{fig6} shows contour plots obtained with the BDF and the CN methods, and figure \ref{fig7} shows the CN solution close to the boundary vs. time: while the global behavior of the solutions is the same, the slow-dying oscillations of the CN method close to the boundary are clearly visible. 

The oscillatory behavior of the CN method is due to the reduced damping of the high-frequency spatial modes (see the formula for $A$, eq. (\ref{ACN})), and it is known to sometime spoil the quality of the CN solutions \cite{bri}. 
These numerical tests indicate that the BDF algorithm -- as an extension of the implicit Euler algorithm --  is a good choice for an integrator in large scale biophysical simulations. 

\section{Conclusions}
\label{final}

We are developing a simulation program, VBL (Virtual Biophysics Lab) where we aim to include a basic description of biochemical and biomechanical features, to simulate cell clusters, and that should eventually be a numerical model of tumor spheroids, a useful and important {\it in vitro} model of solid tumors \cite{suth}: the stakes are high and this is just one of several attempts to use mathematical and physical methods to understand the biochemical and mechanical properties of tumor spheroids \cite{mod1,mod2,mod3,mod4,mod5,mod6,mod7,mod8,mod9,mod10,mod11,mod12,mod13,mod14,mod15,mod16,mod17}. 

\begin{figure}
\begin{center}
\includegraphics[width=10cm]{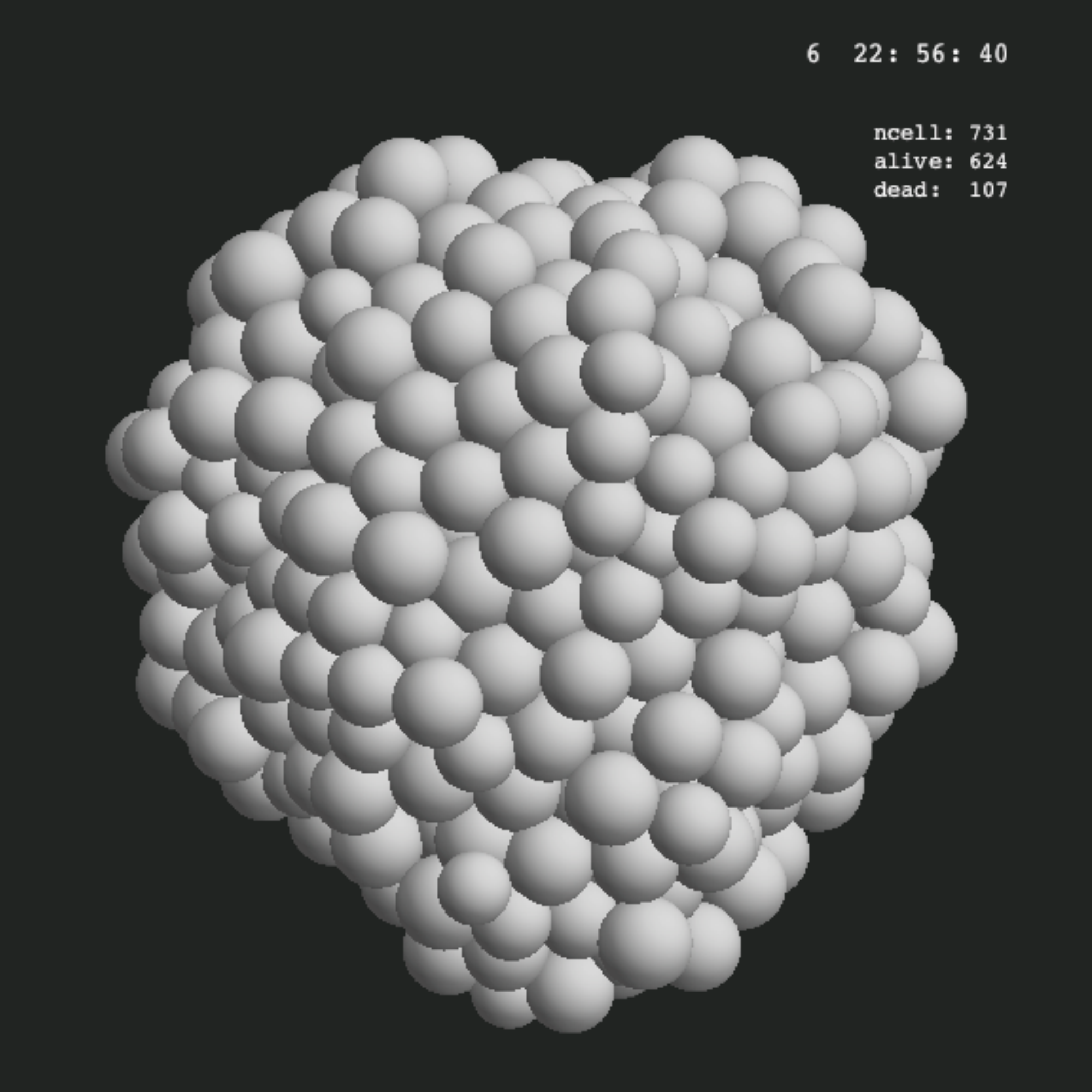}
\caption{\label{fig10} Simulated tumor spheroid in VBL. The simulation started from a single cell and this cluster of cells corresponds to more than 6 days of simulated time.}
\end{center}
\end{figure}

\begin{figure}
\begin{center}
\includegraphics[width=10cm]{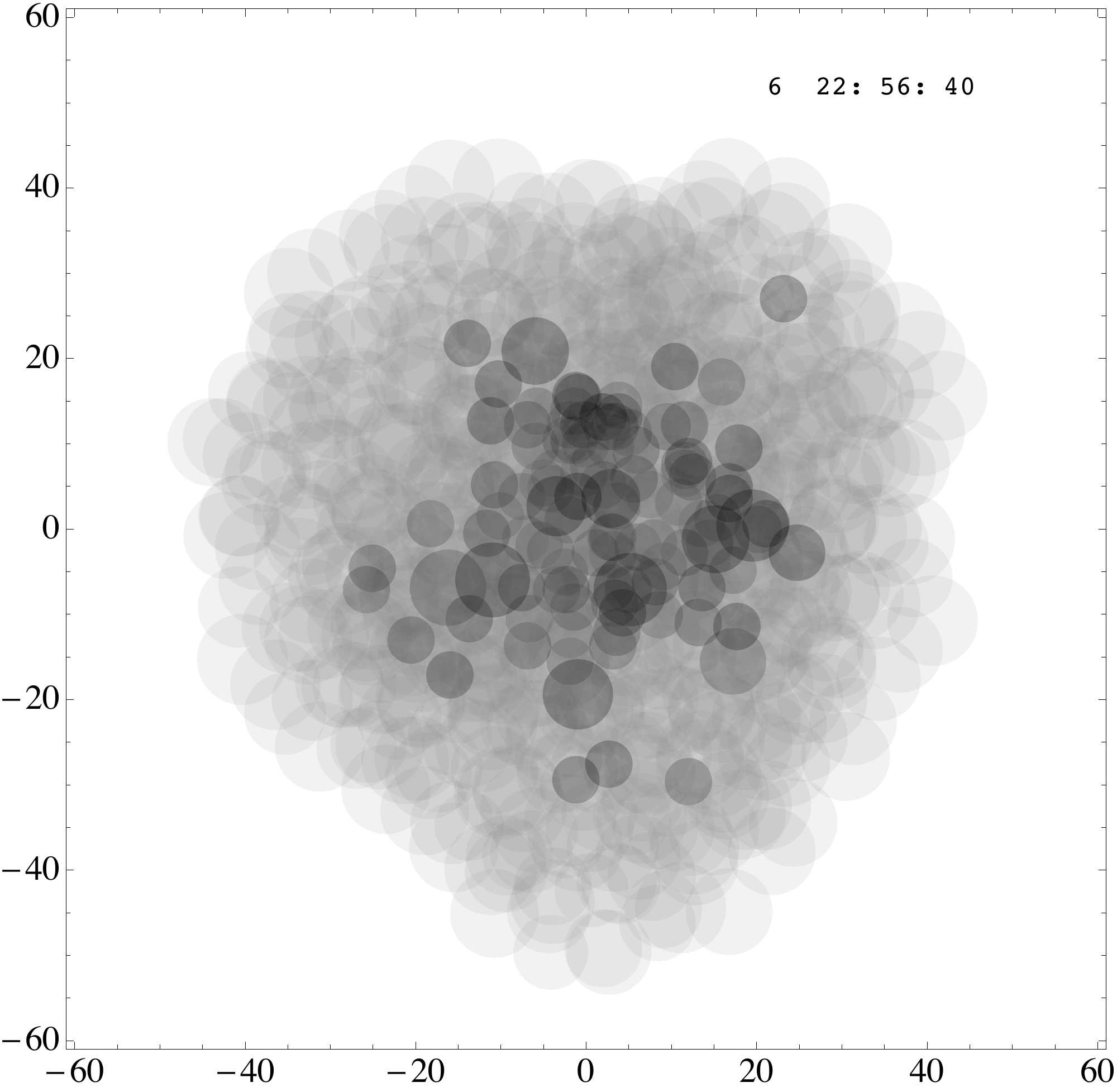}
\caption{\label{fig11} Distribution of dead cells (dark gray) inside the simulated tumor spheroid of figure \ref{fig10}. Here cells are represented as semitransparent balls, and dead cells in the core become visible. The numbers on the axes are $\mu$m.}
\end{center}
\end{figure}

In our modeling effort we proceed in a partly phenomenological way that leads to simple parameterizations: in exchange, we achieve a huge reduction in computational complexity and a considerable reduction of the space-time scale problems that affect simulations aimed at calculating the properties of macroscopic objects starting from microscopic models. We are in an advanced phase of development of the program, and we have already included cell metabolism, growth and proliferation and the extracellular environment. In the previous sections we have discussed the stability properties of the integrators in VBL and in similar programs: we have performed tests on a simple model system and some of the results are interesting on their own, like, e.g., the time evolution and the distribution of G6P in a 1D string of cells (see figures \ref{fig10} and \ref{fig11}).  The 3D part of the program, i.e. geometry and biomechanical interactions, is also included in the prototype version of the program, as well as a description of the extracellular microenvironment. We can  already simulate large populations of dispersed cells, like those in the culture wells used for in vitro growth, and we have produced numerical estimates that are in excellent qualitative agreement, and in good quantitative agreement, with experimental data \cite{VBL1,VBL2,VBL3}. Figure \ref{fig10} shows one simulated spheroid and figure \ref{fig11} is an alternate display that shows the distribution of dead cells inside the same spheroid.

In the present stage of development of the program we are cleaning up the biochemical part of the simulation and building a close integration with the diffusion algorithm to achieve a robust, stable program. Because of discrete -- and sometimes random -- events in the cells' lives, the evolution is only partly described by differential equations. Moreover -- because of cell proliferation -- there is a variable number of equations.  On the basis of the considerations and the numerical tests described in this paper we have decided to settle on the implicit Euler algorithm and its extension, the Backward Differentiation Formula. 
The simulations that produced figures \ref{fig10} and \ref{fig11} were based on much simpler (and unstable) explicit Euler integration steps, and stability problems showed up at an early stage: soon we shall be able to carry out higher quality simulations based on the much more stable and reliable methods described here. 

Eventually we may have to tackle efficiency and speed as the simulations become larger and more time-consuming. However the choice of BDF is open to improvements in speed, since -- as we have seen in the last section -- biochemical changes are comparatively slow, and stiffness is brought about mostly by the diffusion terms in the equations: this means that we may eventually use implicit-explicit methods like those described in \cite{imex} and this may help us retain the advantages of the implicit BDF algorithm and greatly increase speed -- but this is something that we shall study in a future work.



\end{document}